\documentclass{elsart}

\usepackage{natbib}

 \usepackage{epsfig}

\usepackage{amssymb}

\begin{document}

\begin{frontmatter}


\title{The under-explored radio-loudness of quasars and the
 possibility of radio-source--environment interactions}

\author{Katherine M.\ Blundell}
\ead{kmb@astro.ox.ac.uk}
\address{Oxford University Astrophysics, Keble Road, Oxford, OX1 3RH,
  UK}


\author{}

\address{}

\begin{abstract}
I demonstrate that radio observations in the literature to date of
optically-selected quasars are largely inadequate to reveal the full
extent of their jet-activity.  I discuss a recent example of an
optically-powerful quasar, which is radio-quiet according to all the
standard classifications, which Blundell \& Rawlings discovered to
have a $>100$\,kpc jet, and show that other than being the first FR\,I
quasar to be identified, there is no reason to presume it is
exceptional.

I also discuss a possible new probe of accounting for the interactions
of radio sources with their environments.  This tool could help to
avoid over-estimating magnetic fields strengths within cluster
gas.  I briefly describe recent analyses by Rudnick \& Blundell which
confront claims in the literature of cluster gas $B$-fields $> 10
\mu$G.

\end{abstract}




\end{frontmatter}

\section{Two types of quasar --- looks like carelessness?}
\label{sec:twotypes}

The view that there are two distinct types of quasar, differing only
in their radio characteristics, has prevailed for a number of years
(Kellermann et al 1989, Kellermann et al 1994, Miller, Peacock \& Mead
1990, Miller, Rawlings \& Saunders 1993, and Falcke, Sherwood \&
Patnaik 1996).  It now seems hard to avoid the conclusion that the
foundation of these studies, the BQS quasar survey (Schmidt \& Green
1983), has serious and systematic incompletenesses (Goldschmidt et al
1992, Miller et al 1993, Wisotzki et al 2000).  Moreover, four recent
studies suggested that the basis for believing in a bimodality, rather
than a continuity, of the radio properties of quasars may have
evaporated:

[I] Quasars selected from the FIRST survey show no bimodality in
1.4\,GHz-radio luminosity (White et al 2000).  

[II] Quasars from the optically-selected Edinburgh survey show no
bimodality in 5\,GHz-radio luminosity (Goldschmidt et al 1999).  

{\em However, reservations about the lack of a bimodality found by [I]
and [II] include concerns about the high frequency at which the radio
data were obtained, 1.4\,GHz and 5\,GHz respectively: selection and
measurement in a waveband where Doppler boosting is prevalent could
contaminate the measured radio luminosities, and in principle might
blur out any underlying bimodality. }

[III] Consideration of a simple diagram from Blundell \& Rawlings
(2001), reproduced in Figure\,1, shows the potential for the
bimodality perceived in the past to be a consequence of selection in
different radio wavebands.  For example, a standard classification of
radio-quiet quasars (e.g.\ Miller et al 1990) is that a radio-quiet
quasar's luminosity at 5\,GHz is less than $10^{24}\,{\rm
  W\,Hz^{-1}\,sr^{-1}}$.  It is instructive to see how this translates
to classifications concerning FR\,I and FR\,II types at the lower
frequency of 178\,MHz.  Although there have been assertions in the
literature that FR\,I radio structures are never associated with
quasars (Falcke, Gopal-Krishna and Biermann 1995, Baum, Zirbel \&
O'Dea 1995), Figure\,1 suggests that low-frequency extended emission
resembling that well-known around nearby low-power FR\,I radio
galaxies might sensibly be searched for around `radio-quiet' quasars.

[IV] Blundell \& Rawlings (2001) have discovered that one
well-studied, optically-powerful quasar has a 300-kpc FR\,I radio
structure emanating from it.  Its radio luminosity at 5\,GHz falls in
the classification for `radio-quiet' quasars ($10^{23.9}\,{\rm W
Hz^{-1} sr^{-1}}$).  Its radio luminosity at 151\,MHz
($10^{25.3}\,{\rm W Hz^{-1} sr^{-1}}$) is at the transition luminosity
observed to separate FR\,Is and FR\,IIs.

The generality of extended radio emission (presumably from FR\,I-type
[i.e.\ low-power, highly dissipative] jets) in the population of
quasars hitherto deemed `radio-quiet' has yet to be investigated.
Clarification of this issue would considerably benefit studies of the
mechanisms by which quasar central engines work.  To achieve
such aims requires avoiding selecting the quasars in a way which
depends on an unknown and ill-understood mixture of different physical
effects (e.g.\ Doppler boosting, for the FIRST quasars selected at $(1
+ z) \times$ 1.4\,GHz).  It is important to select on {\em
low}-radio-frequency flux-density which should be dominated by lobe or
plume (i.e.\ non-Doppler boosted) jet output.

To establish the prevalence of FR\,I quasars (perhaps this might be a
better way of labelling at least some radio-quiet quasars in the
future), and to deduce their jet-powers, is essential if we are ever
to make definitive deductions concerning the so-called `radio-optical
correlation' (Rawlings \& Saunders 1991) or `jet-disc symbiosis'
(Falcke, Malkan \& Biermann 1995) for quasars and their central
engines.  Only with low frequency data can one hope to do a refined
`radio-optical correlation' analysis and move closer to an
understanding of the relationship between accretion and jet output in
the quasar phenomenon.

\vspace{1.5cm}
\hbox{
\begin{minipage}{3in}
\psfig{figure=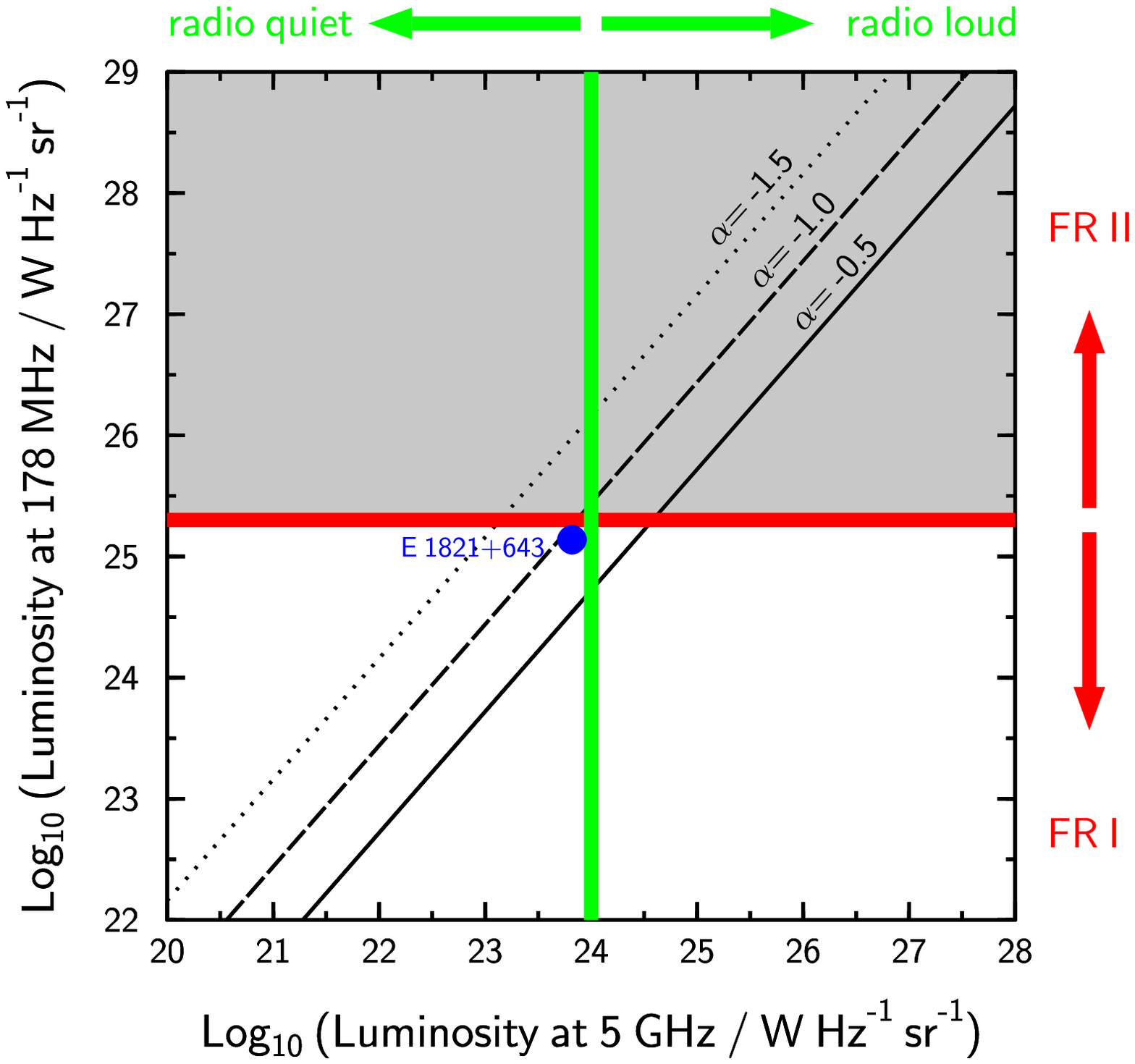,width=3in,angle=0}
\end{minipage}
\hfill
\begin{minipage}{2.35in}
{\bf Figure 1} The vertical green line distinguishes what are
conventionally known as the radio-loud and radio-quiet regimes while
the horizontal red line distinguishes what Fanaroff \& Riley (1974)
observed to separate classical double (FR\,II) and edge-dimmed (FR\,I)
radio sources.  E\,1821+643 is the quasar which was discovered by
Blundell \& Rawlings (2001) to have an FR\,I structure; further
discussion of this plot may be found in that paper.
\end{minipage}
}

\section{A bimodality in the radio observations of quasars?}
\label{sec:bimodality}

Radio-quiet quasars are manifestly not radio-silent quasars, yet it is
ironic that the so-called radio-quiet quasars have historically had
only short snapshot radio observations with typical durations of
minutes (see Figure\,2) while {\em much} deeper observations have been
made of brighter radio-loud quasars.  In other fields of astronomy,
fainter objects are observed for longer integration times than their
brighter counterparts!  Such snapshots are way too short to reach
interesting sensitivity limits (e.g.\ the FR\,I / FR\,II break) for
quasars at even intermediate redshifts.  The first FR\,I radio
structure discovered by Blundell \& Rawlings (2001) was only revealed
following a relatively deep observation ($\sim 2$\,hrs).  Blundell \&
Rawlings showed how this extended structure would simply be missed by
a typical short snapshot observation (their figure 2).

\hbox{
\begin{minipage}{2.7in}
\psfig{figure=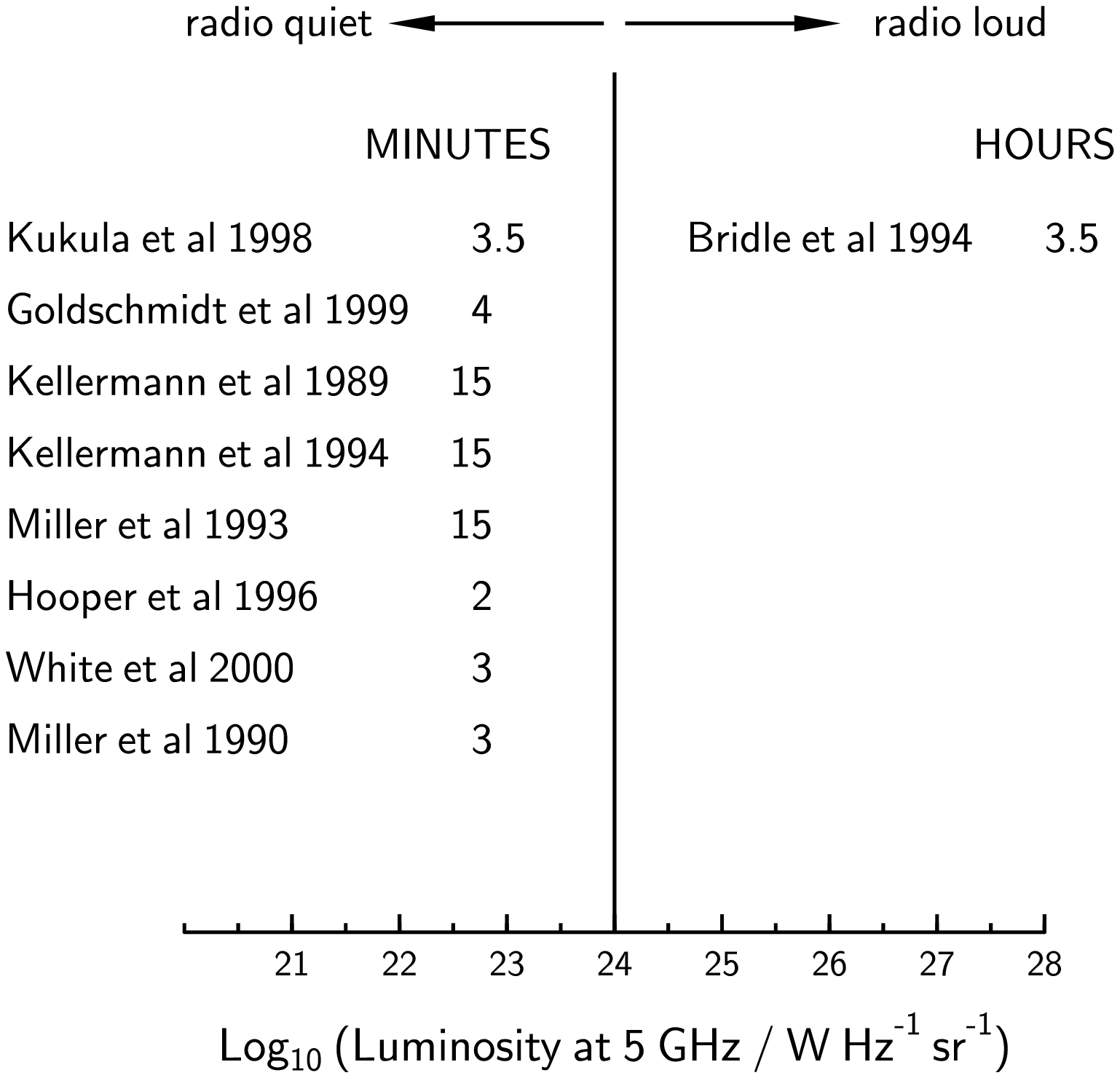,width=2.7in,angle=0}
\end{minipage}
\hfill
\begin{minipage}{2.6in}
{\bf Figure 2} The left column lists the mean length of time-on-source
for observations of radio-quiet quasars in the recent literature; the
units are minutes.  The VLA has been used to make very deep radio
observations (integration times of hours) of radio-loud quasars ---
e.g.\ the study of some 3C quasars by Bridle et al (1994) in which the
radio structure on all scales is fully sampled; comparable quality
observations of quasars which are deemed to be radio-quiet, though
which are actually {\em not} radio-silent, have yet to be made.
\end{minipage}
}

\section{A different picture with long baselines?}
\label{sec:longbaselines}

In addition to significantly shorter integration times, existing
observations of radio-quiet quasars are inferior to those of
radio-loud quasars in another respect: most of the observations listed
in the left column of Figure\,2 have been made with a {\em single}
configuration of the VLA.  Typically this configuration has been one
of the most extended VLA configurations, either A or B.  Observations
in extended configurations are less sensitive to extended structure
(such as that from FR\,I jets for example) than the more compact
configurations as Figure\,3 demonstrates.  On the assumption that the
FR\,I quasar discovered by Blundell \& Rawlings (2001) is not some
lucky exception, then there is the interesting possibility that many
more of the ``radio-quiet quasar'' population have FR\,I radio
structures.  Their recognition is contingent on deep, sensitive
observations at low-frequency, and/or with sufficiently short
baselines that emission on all size scales is actually sampled.

\psfig{figure=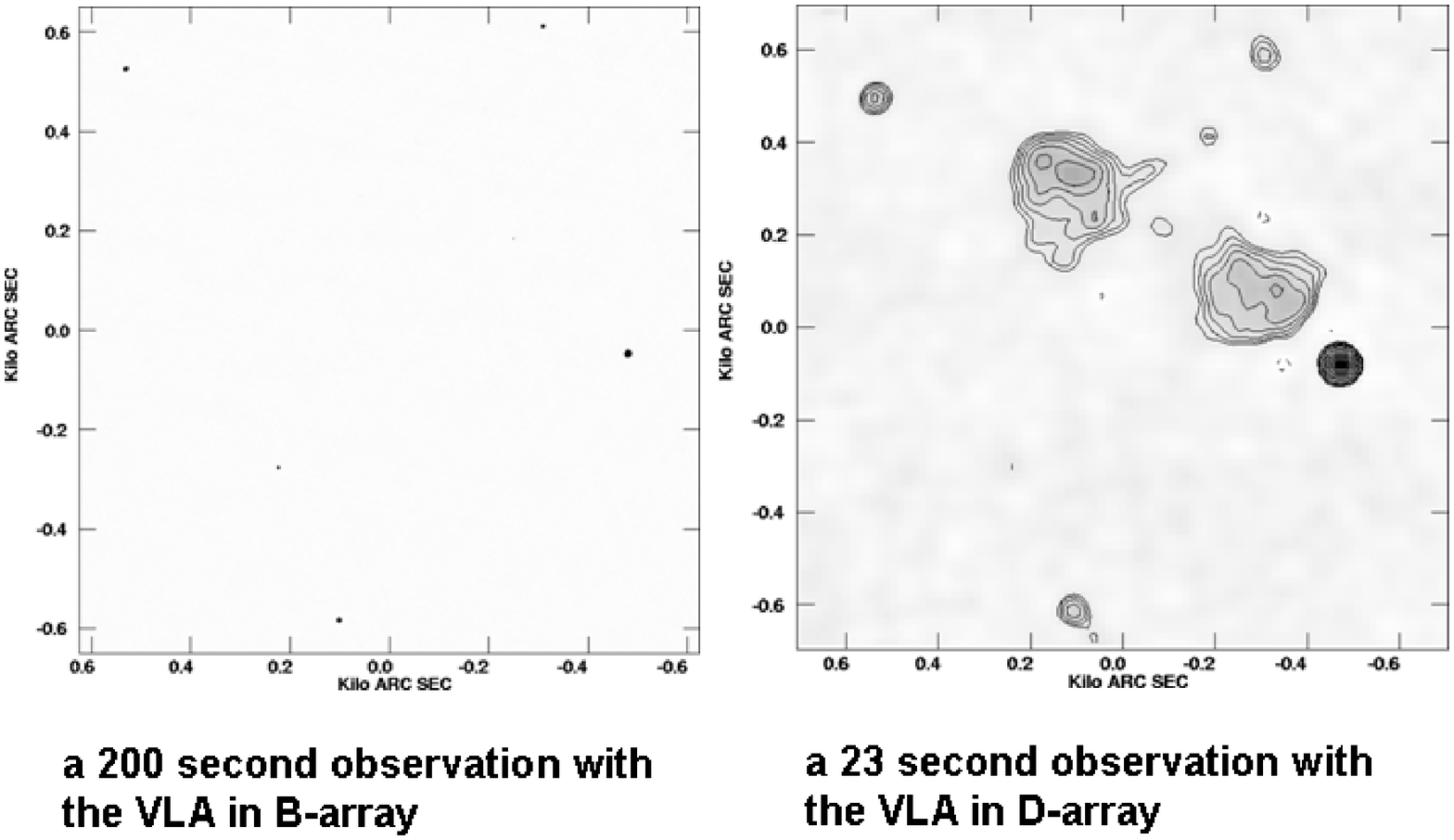,width=\textwidth,angle=0}
\begin{minipage}{\textwidth}
{\bf Figure 3} Images of the same region of sky (left) from the
extended VLA B-configuration [from the FIRST survey] and (right) from
the most compact VLA D-configuration [from the NVSS survey],
demonstrating a basic principle of interferometry that long baselines
are blind to extended structures.
\end{minipage}

\section{A new probe of radio sources interactions with their
  environments? } 
\label{sec:environments}

FR\,I radio sources are well-known for their dissipative jets which
characterize their overall morphology (e.g.\ De~Young 1993).
Entrainment is likely to be a common feature of FR\,I jets.  For
example, Bicknell (1994) has used the conservation laws to demonstrate
the relationship between entrainment and deceleration.  In a similar
manner, Laing \& Bridle (2002) found that mass-loading or entrainment
of ambient matter is required to reproduce the observed deceleration
in the jets of the FR\,I radio galaxy 3C\,31.  Once thermal material
is entrained within the magnetic field of the synchrotron emitting
plasma it must at least be considered as contributing to any observed
Faraday Rotation.  This was the theme of the second part of my talk,
namely: the possible development of a tool to probe entrainment and
interactions local to the radio source, by considering the
relationships between the spatial distributions of Rotation Measure
and of Polarisation Angle across radio sources.  Spatial correlations
of emission line gas and radio lobes (and to a rather lesser extent,
the polarimetric properties of these) have been studied by Heckman
(1981), van~Breugel et al (1984) and Pedelty et al (1989a,b) for
example. 

Rudnick \& Blundell (2003) recently examined the Rotation Measure
distribution of an important test case, PKS\,1246$-$410.  This is a
useful radio source in this regard since, with a physical size of only
10\,kpc, it is entirely embedded within its host galaxy NGC\,4696,
so significant local effects might be expected to be present.  The
rotation measure across this source has nonetheless been used to make
inferences about the magnetic field of the cluster (Centaurus) in
which it resides (Taylor, Fabian \& Allen 2002).

In order to test the plausibility of radio source / environment
interactions manifesting an identifiable signature in the polarimetric
data on this object, Rudnick \& Blundell (2003) looked for
correspondences in the behaviour of the rotation measure distribution
and in the polarisation angle distribution.  They tested out spatial
correspondences in these two distributions, quantifying the extent to
which when one variable changed, the other also changed.  Rudnick \&
Blundell compared these to the results obtained when simulated Rotation
Measure maps were used (whose histograms matched actual data, but had
no spatial information presumed).  Their quantified results indicated
that the correspondences seen in the actual data were significantly
more pronounced than those from the simulated datasets, suggesting
that a signature of local interactions is indeed present in the
Rotation Measure distribution.

The reader is referred to the original paper for details of the
experiments and analysis of this work.  Rudnick \& Blundell (2003)
made a strong case for radio source/environment interactions, but
noted that an artifically constructed unrelated Faraday medium could
possibly mimic the observed correlations.  Indeed, Ensslin et al
(astro-ph/0301552) have proposed just such an artificial distribution,
assuming the same high-order statistical properties for the source and
for the cluster, as a way to rescue the interpretation of RM
variations as due to the overall cluster medium.  If one is to build a
case that the RM variations are due to the cluster medium, then there
should be some test that rules out contributions from the
source/medium interactions (which does not yet exist), as well as
examining models that are physically motivated.  In the meantime,
examination of the role of environmental effects local to the radio
source as a contribution to the observed Rotation Measure urgently
beckons further study, and must be guided by physical models of the
nature of the intra-cluster medium.

\begin{flushleft}
  {\bf Acknowledgements}
\end{flushleft}
I warmly thank collaborators Steve Rawlings and Larry Rudnick, and the
Royal Society for a University Research Fellowship.

\end{document}